\documentclass[12pt,twoside]{article}   %For printing on two sides of page
\usepackage[super,sort,comma]{natbib}
\usepackage{fancyhdr}		%Gives headers and footers defined below
\usepackage[section]{placeins}   %
\usepackage{url}
\usepackage{caption} 
\usepackage{framed,multirow}

\usepackage{amssymb}
\usepackage{amsmath}
\usepackage{latexsym}

\usepackage{color, soul}
\usepackage{xargs}
\usepackage{tikz}
\usepackage{todonotes}
\usepackage{multirow} 
\usepackage{verbatim}
\usepackage{nameref,hyperref}

\usepackage{url}
\usepackage{xcolor}

\usepackage{hyperref}

\usepackage{graphicx}

\makeatletter \renewcommand\@biblabel[1]{$^{#1}$} \makeatother
\renewcommand{\thesection}{{\sf \Roman{section}}.}
\renewcommand{\thesubsection}{\thesection{\sf \Alph{subsection}}.}
\renewcommand{\thesubsubsection}{\thesubsection{\sf \arabic{subsubsection}}.}

\renewcommand{\refname}{}       %

\usepackage{hyperref}
\hypersetup{ colorlinks,
    citecolor=blue,
    filecolor=blue,
    linkcolor=blue,
    urlcolor=blue
}

\usepackage{xcolor}
\definecolor{gray}{rgb}{0.6,0.6,0.6}
\definecolor{red}{rgb}{0.85,0,0}
\definecolor{green}{rgb}{0,0.85,0}
\definecolor{blue}{rgb}{0,0,0.85}
\definecolor{beige}{rgb}{0.92,0.87,0.78}
\usepackage[all]{hypcap}    %causes link to figures to go to figure, not caption
\begin{document}
\title{Tumor aware recurrent inter-patient deformable image registration of computed tomography scans with lung cancer}

\author{Jue Jiang, Chloe Min Seo Choi, Maria Thor, Joseph O. Deasy, Harini Veeraraghavan
\thanks{The manuscript has been submitted on XXX, 2024 for review. This work was partially supported by the NCI R01CA258821 and the Korea Health Technology R\&D Project HI19C1234 and the MSK Cancer Center core grant P30 CA008748. Jue Jiang and Harini Veeraraghavan contributed equally. All authors are with the Department of Medical Physics, Memorial Sloan Kettering Cancer Center, NY, New York, USA.(e-mail: jiangj1@mskcc.org and veerarah@mskcc.org)}}

\maketitle

\begin{abstract}
\noindent {\bf Background:} Voxel-based analysis (VBA) for population level radiotherapy (RT) outcomes modeling requires topology preserving inter-patient deformable image registration (DIR) that preserves tumors on moving images while avoiding unrealistic deformations due to tumors occurring on fixed images. \\
{\bf Purpose:} We developed a \textsc{t}umo\textsc{r}-\textsc{a}ware re\textsc{c}urr\textsc{e}nt \textsc{r}egistration (TRACER) deep learning (DL) method and evaluated its suitability for VBA. \\
{\bf Methods:} TRACER consists of encoder layers implemented with stacked 3D convolutional long short term memory network (3D-CLSTM) followed by decoder and spatial transform layers to compute dense deformation vector field (DVF). Multiple CLSTM steps are used to compute a progressive sequence of deformations. Input conditioning was applied by including tumor segmentations with 3D image pairs as input channels. Bidirectional tumor rigidity, image similarity, and deformation smoothness losses were used to optimize the network in an unsupervised manner. TRACER and multiple DL methods were trained with 204 3D CT image pairs from patients with lung cancers (LC) and evaluated using (a) Dataset I (N = 308 pairs) with DL segmented LCs, (b) Dataset II (N = 765 pairs) with manually delineated LCs, and (c) Dataset III with 42 LC patients treated with RT. \\
{\bf Results:} TRACER accurately aligned normal tissues. It best preserved tumors, \textcolor{black}{indicated by the smallest tumor volume difference} of 0.24\%, 0.40\%, and 0.13 \% and \textcolor{black}{mean square error in CT intensities of 0.005, 0.005, 0.004, computed between original and resampled moving image tumors}, for Datasets I, II, and III, respectively. It resulted in the smallest planned RT tumor dose difference computed between original and resampled moving images of 0.01 Gy and 0.013 Gy when using a female and a male reference. \\
{\bf Conclusions:} TRACER is a suitable method for inter-patient registration involving lung cancers occurring in both fixed and moving images  \textcolor{black}{and applicable to voxel-based analysis methods}. 
\end{abstract}

\section{Introduction}
\begin{figure*}[htp]
\begin{center}		
    \includegraphics[width=1\columnwidth,scale=1]{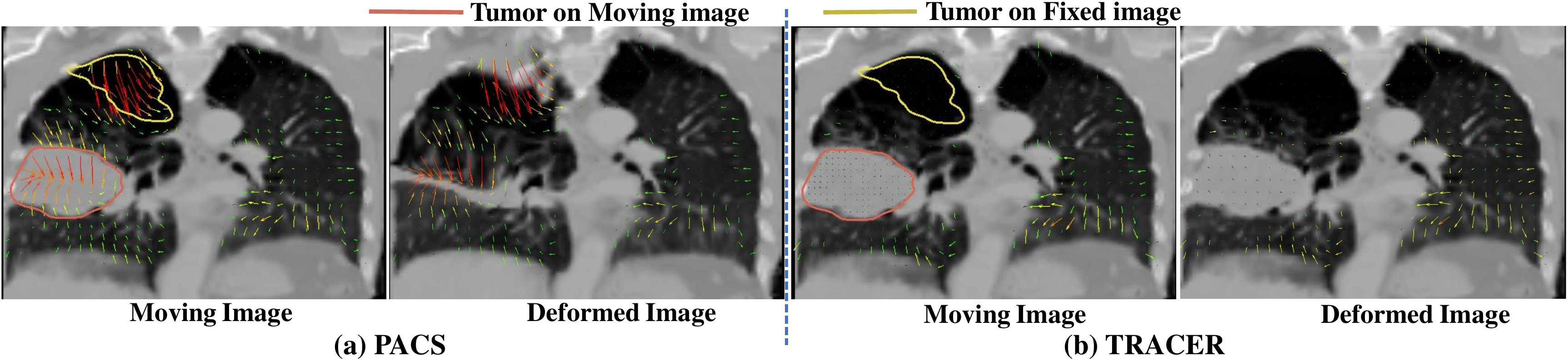}
    \vspace{-0.05cm}\setlength{\belowcaptionskip}{-0.4cm}\setlength{\abovecaptionskip}{0.08cm}\caption{\small Rationale for tumor conditioned \textcolor{black}{inter-patient} registration. (a) Lack of tumor conditioning in PACS, ill-preserves tumors and produces physically unrealistic stretching of tissues. (b) Tumor conditioning in TRACER preserves tumor and normal tissue topology.} \label{fig:rationale}
\end{center}
\end{figure*}

Technological advances in image guidance and radiotherapy (RT) has resulted in several new available treatment options. 
Voxel-based analysis (VBA) is a useful technique for studying population-level outcomes in radiation therapy (RT), such as radiation-induced toxicity. It involves comparing imaging data at the voxel level \textcolor{black}{between different patients, and provides} detailed quantification of how different regions within an organ respond to spillover radiation \textcolor{black}{at a population level}\cite{palma2016,Craddock2023}. However, VBA requires topology preserving deformable image registration (DIR) \textcolor{black}{between different patient scans, or inter-patient DIR, so that voxels from different patients are brought to a common reference coordinate frame for comparing voxel-level dose sensitivities between patients. Accurate inter-patient DIR is thus essential for reliably accurate identification of dose-sensitive regions within organs using VBA}. 

VBA methods typically use iterative B-spline DIR\cite{Craddock2023, palma2016,Beasley2019} to align static 3D scans of different patients to a reference anatomy, because these methods do not require disease specific training. \textcolor{black}{Aligning different patient scans is challenging due to anatomy differences, wherein tumor present in one patient has no physiological or spatial correspondence with tumor occurring in another patient. If care is not taken to handle anatomically different image regions occurring in the moving (patient scan from studied population) and fixed image (reference scan)}, inaccurate registrations can occur, leading to dose response variability\cite{monti2018}, and exclusion of inaccurately aligned patient scans for VBA\cite{MYLONA2019,Beaumont2019}. VBA performed with \textcolor{black}{less accurate inter-patient DIR method} may not be representative of the broader population of patients with the studied disease, \textcolor{black}{motivating our approach}. We developed a novel deep learning (DL) DIR method to align static 3D computed tomography (CT) images of entirely different patients called \textsc{t}umo\textsc{r}-\textsc{a}ware re\textsc{c}urr\textsc{e}nt \textsc{r}egistration (TRACER). TRACER handles scenarios where lung tumors occur in non-corresponding spatial locations in the moving and fixed images. We focus on aligning static 3D instances of one patient anatomy to a reference patient anatomy. 

Inter-patient DIR has been studied extensively in the context of atlas based registration\cite{dalca2019unsupervised,xu2019deepatlas,haq2019} for normal tissue segmentation. DL-DIR methods have typically focused on aligning patient scans without visible tumors\cite{liu2021same,chen2022transmorph}. Although normal anatomy varies between patients, organs occupy fixed anatomic locations unlike tumors. Hence, presence of tumors adds additional complexity in achieving reliably accurate inter-patient registration, necessitating careful selection of the canonical patient for VBA. Groupwise registration methods can overcome the need for careful selection of canonical patient by aligning all patients in a group to an average reference\cite{Ahmad2019,Zhang2021}. However, computational and memory requirements of this approach is prohibitive as all images in the group are required as inputs to the model.

Rigidity constraints are commonly used in DIR to preserve geometry of both normal tissues and tumors\cite{staring2007,Konig2016RadOnc} applied to intra-patient longitudinal registration, with an explicit assumption that no new tumors occur on fixed images. Most deep learning methods performing single step registration cannot preserve topology due to their assumption of a small deformation framework that is approximated using a stationary velocity field (SVF)\cite{Mok2020}. Inverse consistency regularization using cyclical image registration\cite{Kim2021CycleMorph}, inverse flow generation constraints\cite{ZhangInverseConsistent2018}, as well as bidirectional registration to an intermediate space\cite{Mok2020} are some example methods used to ensure topologically consistent deformations. However, such methods focus on achieving flow path invariant deformations, which doesn't ensure preservation of structures of interest. Similarly, recurrent networks\cite{Sandkuhler2019,jiang2022CBCTTMI} can handle large anatomy differences by performing progressive deformation refinement but cannot preserve tumors and cause physically unrealistic tissue stretching in locations containing tumors on fixed images without tumor conditioning (Figure.~\ref{fig:rationale}(a)). 

TRACER builds on our prior work called patient specific anatomic and shape context (PACS)\cite{jiang2022CBCTTMI} for aligning intra-patient CT with during treatment cone beam CTs. Similar to PACS, it employs a 3D image registration network where the encoder layers consist of stacked 3D convolutional long short term memory network (3DCLSTM) used to progressively refine the extracted features and ultimately the deformations. TRACER differs from PACS by \textcolor{black}{extending the use case to inter-patient DIR for image volumes with non-corresponding tumors. It introduces input tumor conditioning combined with bi-directional tumor rigidity implemented to preserve tumors in the forward flow and penalize distortions within tumor voxels occurring in fixed images in the inverse flow.} Input tumor conditioning also relaxes rigidity penalty by allowing tissues adjacent to the conditioned structure to deform. Our approach can be fully automated by using auto-segmented tumors for conditioning the registration. TRACER thus addresses the shortcommings of prior methods and effectively preserves tumors while also avoiding unrealistic tissue stretching (Figure.~\ref{fig:rationale} (b)). To our best knowledge, this is the first DL-DIR work to align different patient scans containing tumors in non-corresponding locations on moving and fixed images. 

Our contributions are: (a) a new inter-patient recurrent DIR method that is robust to tumors occurring in both fixed and moving images, (b) tumor conditioned DIR to preserve \textcolor{black}{geometry and CT intensity of tumors} and prevent unrealistic distortions due to non-corresponding tumors occurring on fixed images, (c) fully automated approach using deep learning tumor segmentation to condition DIR, and (d) rigorous testing of our approach on three datasets in patients with locally advanced non-small cell lung cancer (LA-NSCLC) treated with RT.

\begin{figure}[t]
		\begin{center}
			\includegraphics[width=1.0\columnwidth,scale=1]{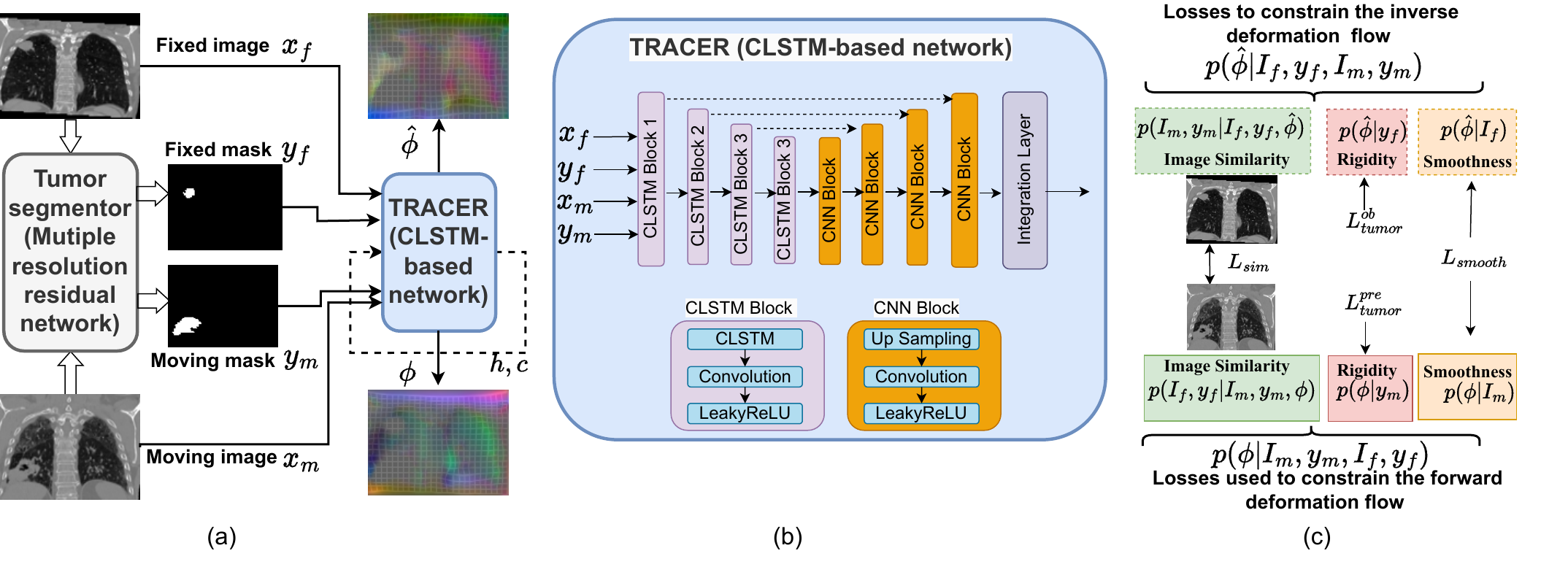}
			\vspace{-0.05cm}\setlength{\belowcaptionskip}{-0.4cm}\setlength{\abovecaptionskip}{0.08cm}\caption{\small (a) Schematic of tumor aware recurrent inter-patient DIR (TRACER) showing the fixed image and its mask ($\{x_{f}, y_{f}\}$), moving image and its mask ($\{x_{m}, y_{m}\}$ and the hidden layer $h$ input to the TRACER. Tumor conditioning is automated using published method. (b) shows TRACER architecture. (c) shows the losses used to optimize the network.} \label{fig:method}
		\end{center}
\end{figure}

\section{Method }
\textbf{Problem setting and approach: \/}\rm Given a pair of rigidly aligned static 3D CT scans from \textcolor{black}{two different patients} $\{I_m, I_f\}$, the goal is to deformably align moving $I_{m}$ to fixed image $I_{f}$ while preserving topology and all visible tumors occurring on moving image. \textcolor{black}{This work is concerned with aligning two different and unrelated patient scans where tumors occurring in one patient or moving image $I_m$ have no relation to the tumors occurring in a different patient or fixed image $I_f$}. We achieve this goal by conditioning the registration using the tumor segmentation masks $\{y_{m}, y_{f}\}$ as additional input channels with $\{I_{m}, I_{f}\}$ to the model (Figure.~\ref{fig:method}). Input structure conditioning modulates the extracted image features, and tumor masks are also used when computing image similarity and bidirectional rigidity penalties. 

\subsection{Tumor-aware recurrent registration network (TRACER):}
TRACER is a structure-conditioned registration method, which combines 3D input image pairs with corresponding 3D structure segmentation masks ($\{ I_{m}, I_{f}, y_{m}, y_{f} \}$) as its inputs. Our approach uses automated tumor segmentations produced using a published multiple resolution residual network (MRRN), available through open-source CERR software\footnote{ https://www.github.com/cerr/CERR} (Figure.~\ref{fig:method}(a)). TRACER architecture (Figure.~\ref{fig:method}(b)) uses a 3D-Unet style architecture where each encoder layer consists of a 3DCLSTM layer followed by a convolutional layer with stride of 2 followed by LeakyReLU activation. In addition to progressively refining the extracted deformations, stacked 3D-CLSTM architecture allows continuity of computed features between the various feature levels. 

The moving image ($I_{m}$), fixed image($I_{f}$), and the segmentations ($y_{m}$ and $y_{f}$) are input to the first CLSTM step at $t=1$, with the hidden state initialized with zero. Subsequent CLSTM steps use the output consisting of the deformed moving image and the corresponding resampled segmentation and the fixed image and its segmentation as well as the hidden state updated from prior CLSTM step as their inputs (Figure.~\ref{fig:method} (a)). Structure conditioning is explicitly implemented into the CLSTM layers by updating the gating filters consisting of forget gate $f^{t}$, memory cells $c^{t}$, hidden state $h^{t}$, input state $i^{t}$, and output gate $o^{t}$ as below:
\begin{equation}
\begin{split}
f^{t}&=\sigma(W_{xf}*\overbrace{\{ I_{m}^{t-1}, y_{m}^{t-1}, I_{f}, y_{f} \}}^{\text {Input State for $x^{t}$}}+W_{hf}*h^{t-1}+b_{f})\\
i^{t}&=\sigma(W_{xi}*\{ I_{m}^{t-1}, y_{m}^{t-1}, I_{f}, y_{f} \}+W_{hi}*h^{t-1}+b_{i})\\
\tilde{c}^{t}&=\textcolor{black}{\mathrm{tanh}}(W_{x\tilde{c}}*\{ I_{m}^{t-1}, y_{m}^{t-1}, I_{f}, y_{f} \}+W_{h\tilde{c}}*h^{t-1}+b_{\tilde{c}})\\
o^{t} &= \sigma(W_{xo}*\{ I_{m}^{t-1}, y_{m}^{t-1}, I_{f}, y_{f} \}+W_{ho}*h^{t-1}+b_{o})\\
c^{t} &= f^{t} \odot c^{t-1}+i^{t}\odot \tilde{c}^t\\
h^{t} &= o^{t} \odot \textcolor{black}{\mathrm{tanh}}(c^{t}),
\end{split}
\end{equation}
where, $\sigma$ is the sigmoid activation function, $*$ the convolution operator, $\odot$ the Hadamard product, and $W$ the weight matrix. 

The decoder layers (same number as the encoder layers) of the registration network are composed of convolutional filters with up-sampling operations to extract the dense 3D flow field. Skip connections are implemented between the encoder and decoder layers to improve training stability and the sharpness of extracted deformations. The flow field is converted into deformation vector field (DVF) using spatial transform networks\cite{jaderberg2015spatial}. Fast computation is achieved by assuming a stationary velocity field (SVF) $v_{j}$ for the deformation computed at the end of each CLSTM step. Hence the DVF $\phi^{j}$ in each step $j$ is computed as $\frac{d\phi_{i}^{j}}{dj} = v^{j}(\phi_{i}^{j})$, subject to $\phi_{i=0}^{j} = Id$ using spatial time integration ($i=7$)\cite{Ashburner2007,Mok2020,balakrishnan2019voxelmorph} performed using scaling and squaring transforms\cite{dalca2019unsupervised}. Diffeomorphic integration layer is implemented to convert the incremental flow field into DVF while ensuring diffeomorphic deformation\cite{dalca2019unsupervised}. The velocity $v_{j}$ is allowed to vary for each CLSTM step, which in turn allows computation of a temporally and spatially varying displacement field. Note that our computation using a SVF framework is performed for the incremental deformation computed for each CLSTM step as opposed to methods that use SVF for computing the entire deformation in a single step\cite{balakrishnan2019voxelmorph,dalca2019unsupervised}. Our approach thus ensures higher accuracy in the presence of large deformations. The progressive sequence of deformations to deform the moving image $x_m$ to fixed image $x_f$ in $T$ CLSTM steps, with each CLSTM step using $N$ time integrations is thus computed as, $\phi_{m}^{f} = \circ_{t=0}^{T} \circ_{i=0}^{N} \phi_{i}^{t}(v^{t})$, where $\circ_{t=0}^{T} \phi^{t}(x_{m}) = \phi^{1}  \circ (\phi^{2} (... \circ (\phi^{T}, x_{m})))$ is a composition operator.

\subsection{Tumor-aware network regularization}
The network is optimized by maximizing the posterior probability $P(\phi|I_{f},y_{f},I_{m}, y_{f})$ of the forward dense deformation flow used to map the moving image (\textcolor{black}{$I_{m}$}) into the spatial coordinates of the fixed image (\textcolor{black}{$I_{f}$}).
\textcolor{black}{The posterior $P(\phi|I_{f},y_{f},I_{m}, y_{m})$ can be decomposed as: 
\begin{equation}
\setlength{\abovedisplayskip}{1pt}
\setlength{\belowdisplayskip}{1pt} 
\begin{split}
& \mathrm{log} P(\phi | I_{m}, y_{m}, I_{f}, y_{f}) = \\ & 
\mathrm{log} P(I_{f},y_{f} | I_{m}, y_{m}, \phi) + \mathrm{log} P(\phi | I_{m}, y_{m}) + \mathrm{log} P(y_{m} | I_{m}) -\mathrm{log} P(I_{f},y_{f} | I_{m}, y_{m}) - \mathrm{log} P (y_{m}|I_{m})
\label{eqn:forward_full}
\end{split}
\end{equation}
The terms ($I_{f},y_{f}$) and ($I_{m}, y_{m}$) are independent and they come from two different patients that have no temporal or sequential relationship\cite{yang2018conditional}. Hence, Eqn.~\ref{eqn:forward_full} can be simplified as:
\begin{equation}
\setlength{\abovedisplayskip}{1pt}
\setlength{\belowdisplayskip}{1pt} 
\begin{split}
& \approx \mathrm{log} P(I_{f},y_{f} | I_{m}, y_{m}, \phi) + \mathrm{log} P(\phi | I_{m}, y_{m}) \\
& \approx  \mathrm{log} P(I_{f},y_{f} | I_{m}, y_{m}, \phi) + \mathrm{log} P(\phi | I_{m}) + \mathrm{log} P (\phi | y_{m}) \hspace{3pt} + \mathrm{log} P(y_{m} | I_{m}). 
\end{split}
\end{equation}
The term $P(y_{m}|I_{m})$ is constant because the segmentation is computed prior to optimization using the DIR network. The posterior $P(\phi|I_{f},y_{f},I_{m}, y_{m})$ can now be expressed as: 
\begin{equation}
\setlength{\abovedisplayskip}{1pt}
\setlength{\belowdisplayskip}{1pt} 
\begin{split}
\textcolor{black}{\mathrm{log}}P(\phi|I_{f},y_{f},I_{m}, y_{m}) & = \underbrace{\mathrm{log} P(I_{f},y_{f} | I_{m}, y_{m}, \phi)}_\text{Similarity} + \underbrace{\mathrm{log} P(\phi | I_{m})}_\text{Smoothness} + \underbrace{\mathrm{log} P (\phi | y_{m})}_\text{Rigidity} \textcolor{black}{+ \mathrm{constant}}. \\
& =L_{sim} + L_{smooth} + L_{tumor}^{pre}
\label{eqn:forward}
\end{split}
\end{equation}
}

Image similarity ($L_{sim}$) in Eqn.~\ref{eqn:forward} is computed by removing the effect of tumors in moving and fixed image as: 
\begin{equation}
\setlength{\abovedisplayskip}{1pt}
\setlength{\belowdisplayskip}{1pt} 
\begin{split}
    L_{sim} = \underset{t=1}{\overset{N}\sum}\ E|| I_{f} \odot (I - y_{f}), I_{m}^{t} \odot (I - y_{m}^{t})|| /N,  
\end{split} 
\label{eqn:similarity}
\end{equation} where $I$ is the identity matrix. 
\\
\textcolor{black}{The smoothness term $P(\phi|I_{m})$ or $L_{smooth}$ in Eqn.~\ref{eqn:forward} is computed over all voxels $p \in \Omega$ in the resampled $I_{m}$, where $p$ is 1 when it is within the dimensions of $I_m$ and 0 everywhere else. This term regularizes the deformations\cite{MokChung2021MICCAI} and is computed by uniformly sampling the deformation field $\phi^{t}$ in $\Omega$ at each CLSTM step $t$} as, 
\begin{equation}
\setlength{\abovedisplayskip}{1pt}
\setlength{\belowdisplayskip}{1pt} 
\begin{split}
L_{smooth} = \underset{t=1}{\overset{N}\sum}\ \underset{p \in \Omega}{\sum}\ ||\nabla \phi^t(p)||^{2}/N. 
\end{split} 
\label{eqn:smoothness}
\end{equation}
Finally, rigidity or tumor preservation loss $L_{tumor}^{pre}$ in Eqn.~\ref{eqn:forward} penalizes any forward deformation flow within tumor voxels in moving image by requiring the Jacobian determinant $J(y)=det(\nabla\phi(y))$ to be equal to 1, which corresponds to no mass change\cite{niedzielski2016}. Of note, this loss only constrains the deformation within the tumor voxels such that voxels outside the tumor geometry are allowed to deform. This loss is computed in N CLSTM steps as:
\begin{equation}
\setlength{\abovedisplayskip}{1pt}
\setlength{\belowdisplayskip}{1pt} 
\begin{split}
L_{tumor}^{pre} = \underset{t=1}{\overset{N}\sum}L_{tumor}^{pre,t}=\underset{t=1}{\overset{N}\sum}\ |\mathrm{det}(y^{t-1}_{m}\cdot \nabla\phi^t)-1|_2/N.
\end{split}
\label{eqn:forward_volume_def}
\end{equation}
\\
The formulation in Eqn.~\ref{eqn:forward} is sufficient to achieve topology preserving deformations when no tumors are present in the fixed image. In order to account for any tumors occurring on the fixed image, similar to \ref{eqn:forward}, we constrain the inverse deformation flow $\hat{\phi}$ as: 
\begin{equation}
\setlength{\abovedisplayskip}{1pt}
\setlength{\belowdisplayskip}{1pt} 
\begin{split}
    \textcolor{black}{\mathrm{log}} P(\hat{\phi} | I_{m}, y_{m}, I_{f}, y_{f}) = & \underbrace{\textcolor{black}{\mathrm{log}} P(I_{m},y_{m} | I_{f}, y_{f}, \hat{\phi})}_\text{Similarity} + \underbrace{\textcolor{black}{\mathrm{log}} P(\hat{\phi} | I_{f})}_\text{Smoothness} + \underbrace{\textcolor{black}{\mathrm{log}} P (\hat{\phi} | y_{f})}_\text{Rigidity}  \textcolor{black}{+ \mathrm{constant}}. \\
    & \textcolor{black}{=\hat{L}_{sim} + \hat{L}_{smooth} + L_{tumor}^{ob}}
\label{eqn:inverse}
\end{split}
\end{equation} 
The image similarity in Eqn.~\ref{eqn:inverse} is computed similar to Eqn.~\ref{eqn:similarity} using the resampled fixed image $I_{f}^{t}$ and segmentation $y_{f}^{t}$ at each CLSTM step $t$ into the spatial coordinates of the moving image using $\hat{\phi^{t}}$. $\hat{\phi^{t}}$ is computed by the integration of negative flow field $\hat{\phi^{t}} = -\phi^{t}$ such that $\phi^{t} \circ \hat{\phi^{t}}$ = $\textcolor{black}{\mathrm{exp}}(t) \circ \textcolor{black}{\mathrm{exp}}(-t)$ = $\textcolor{black}{\mathrm{exp}}(t - t)$ = $Id$\cite{dalca2019unsupervised}. The approximation of inverse flow field as $-\phi$ is possible due to the small displacements resulting from a progressive deformation framework combined with a time integration performed inside each CLSTM\cite{Ashburner2007}. Smoothness loss is computed in the same way as in the forward case but with inverse DVF. 

The rigidity penalty in Eqn.~\ref{eqn:inverse} is applied to tumor voxels occurring in the fixed image. This penalty is called tumor obliteration loss $L_{tumor}^{ob}$ and it forces non-deformation of voxels on moving image that map to the voxels within the tumors occurring in the fixed image. The same loss is computed in 1 to $N$ CLSTM steps as: 
\begin{equation}
\setlength{\abovedisplayskip}{1pt}
\setlength{\belowdisplayskip}{1pt} 
\begin{split}
L_{tumor}^{ob} = \underset{t=1}{\overset{N}\sum}L_{tumor}^{ob,t}\underset{t=1}{\overset{N}\sum}\ |\textcolor{black}{\mathrm{det}}(y^{t-1}_{f}\cdot \nabla\hat{\phi^t})-1|_2/N,
\end{split}
\label{eqn:inverse_volume_def}
\end{equation}
The total loss is computed as, \textcolor{black}{$L_{reg}$ = $L_{sim}$ + $\hat{L}_{sim}$+ $\lambda_{smooth}$ ($L_{smooth}$+$\hat{L}_{smooth}$) + $\lambda_{pre}$ $L_{tumor}^{pre}$+ $\lambda_{ob}$ $L_{tumor}^{ob}$}. \textcolor{black}{$L_{sim}$ and $\hat{L}_{sim}$ are the image similarity loss; $L_{smooth}$ and $\hat{L}_{smooth}$ are the smoothness loss and $L_{tumor}^{pre}$ and $L_{tumor}^{ob}$ are the tumor rigid constraint.}

\subsection{Training \textcolor{black}{dataset} and details}
TRACER as well as compared DL-DIR methods were optimized in an unsupervised manner without any ground truth DVFs using identical training images and implemented using the Pytorch library to align 3DCTs from different patients. Contrast enhanced 3DCT image pairs (n = 32,220) from 180 patients with locally advanced non-small cell lung cancer (LA-NSCLC) treated with intensity modulated RT at our institution were used to train the models and then validated with 552 CT pairs from 24 patients not used in training to select the models for testing. All networks were trained on Nvidia A100 with 80GB memory and optimized using ADAM algorithm with an initial learning rate of 0.0002 for the first 50 epochs and then decayed to 0 in the next 50 epochs and a batch size of 4. We set $\lambda_{smooth}$=25, $\lambda_{pre}$ =1000 and $\lambda_{ob}$=1000 experimentally. Both TRACER and PACS used 8 3D-CLSTM steps. All models were "locked" for testing performed on entirely different datasets as described below and tuned to the best of our ability in order to achieve the best performance possible. 
\\
\textbf{Preprocessing details:\/}\rm All CT scans were cropped to contain the chest region with 10 cm margin added to include a small portion of air outside the body such that both lungs and the chest enclosing the lungs were visible in all scans and then resampled to  128$\times$ 128 $\times$ 96 volume. Images were resampled back to the original volumes for accuracy calculation.

\subsection{Testing datasets}
\textbf{Dataset I\/} \rm consisting of 380 image pairs (20 $\times$ 19) sourced from a public dataset\cite{kalendralis2020fair} of pre-RT thoracic CT scans of LA-NSCLC patients (tumor size between 4.88 cc to 356 cc) was used to evaluate registration accuracy when using auto-segmented tumors as conditioning inputs. Expert tumor delineations were available and used for accuracy evaluation.  
\\
\textbf{Dataset II \/}\rm consisting of 756 CT image pairs (28 $\times$ 27) from a randomly selected set of 28 LA-NSCLC patients with expert delineations from public dataset\cite{aerts2015data} and treated with definitive IMRT were evaluated. Tumors ranged in size between 0.022 cc to 640 cc and were distributed across the various lobes of the lungs including the mediastinum. 
\\		 
\textbf{Dataset III \/} \rm consisting of 42 instituitional LA-NSCLC patients (23 male and 19 female) treated with IMRT to 60 Gy in 2 Gy fractions were aligned to a canonical reference to study utility of TRACER for VBA. Retrospective analysis was approved by the local institutional review board at the Memorial Sloan Kettering Cancer Center.

\subsection{Metrics and statistical analysis}
Registration accuracy was evaluated in terms of healthy tissue segmentation accuracy computed using the Dice similarity coefficient (DSC) \textcolor{black}{and} Hausdroff distance at 95th percentile (HD95) metrics for the lungs, heart, and spinal cord by comparing against deep learning model\cite{um2020multiple} used to generate clinical segmentations. Instead of measuring target registration error, which requires identification of matching landmarks that is hard to locate on two different patient anatomies, we measured the distance between the medial axes of tubular structures including, the pulmonary artery (PA), aorta, inferior venacava (IVC), and trachea from a subset of 26 patients with available delineations. This distance, called \textbf{m}edian of \textbf{c}losest points \textbf{d}istance (MCD) measured misalignments along the entire organ length. 

Accuracy of tumor preservation was computed using tumor volume loss or $\Delta T = \frac{|V_{def}-V_{m}|}{V_{m}}\times100$, where $V_{m}$, $V_{def}$ are the volume of tumors in the original and deformed moving images, respectively. The percentage mean \textcolor{black}{\textbf{l}}ocal \textcolor{black}{\textbf{ex}}pansion and \textcolor{black}{\textbf{s}}hrinkage inside the tumor in the moving image (M$_{lexs}$\%) was calculated using the Jacobian determinant as $\frac{\Sigma \|J_{i}-1\| \cdot y_{m}}{N} \times 100$\cite{ALAM2021883}, where $N$ is the number of voxels within the tumor mask in the original moving image. \textcolor{black}{We also evaluated the tumor preservation by computing the mean squared error (MSE) between the resampled tumor mask within the resampled moving image and the original tumor mask within the moving image.} In a subset of 42 patients (Dataset III) who had radiation dose maps available, we also computed difference in the Planned Tumor Dose ($\Delta$PTD) between undeformed and deformed moving images. \textcolor{black}{Registration accuracy was computed by comparing the resampled lung contour from moving patient with the lung contour of the reference patient using DSC overlap measure. A DSC $<$ 0.8 was deemed as poor registration. The number of patients with poor registration were calculated.} 
Statistical comparisons of TRACER against the various methods was performed using two-sided, paired Wilcoxon signed-rank tests at 95\% confidence interval. 

\section{Experiments configuration}
\subsection{Comparative experiments} TRACER (1,026,771 parameters) was compared against PACS\cite{jiang2022CBCTTMI} (522,723 parameters) model as the two models share similar encoder-decoder structure. It was also compared against fast symmetric registration (FastSyM)\cite{Mok2020}, which performs bidirectional alignment to compute diffeomorphic registration, transformer based Transmorph\cite{chen2022transmorph}, and iterative DIR called SyN\cite{avants2008symmetric}. 

\subsection{Ablation experiments}
Ablation experiments were done using \textcolor{black}{Dataset II}. Ablation experiments studied the impact of various losses and CLSTM on tumor volume preservation and normal tissue segmentation accuracy. Impact of number of CLSTM layers was also analyzed in the context of strength of encoder feature activation and segmentation accuracy of the heart as it occurs centrally.   

\subsection{Results}
\subsubsection{Registration accuracy\/}\rm
Table~\ref{tab:reg_acc_stats} shows the tumor preservation metrics computed from the three datasets. TRACER best preserved tumor volumes as shown by the lowest tumor loss $\Delta$T and shrinkage $M_{lexs}$ metrics in all testing datasets. \textcolor{black}{It also resulted in the least MSE indicating that the CT intensities within the moving tumor image were preserved upon deformation.} PACS on the other hand was least effective in preserving tumor, when using local tumor expansion and shrinkage measure. \textcolor{black}{PACS was similarly accurate as the Transmorph method when using MSE measure.} Both FastSym\cite{Mok2020} and transformer-based Transmorph\cite{chen2022transmorph} were less accurate than TRACER, indicating the importance of tumor conditioning.

\begin{figure}[h]
		\begin{center}
			\includegraphics[width=0.8\columnwidth,scale=1]{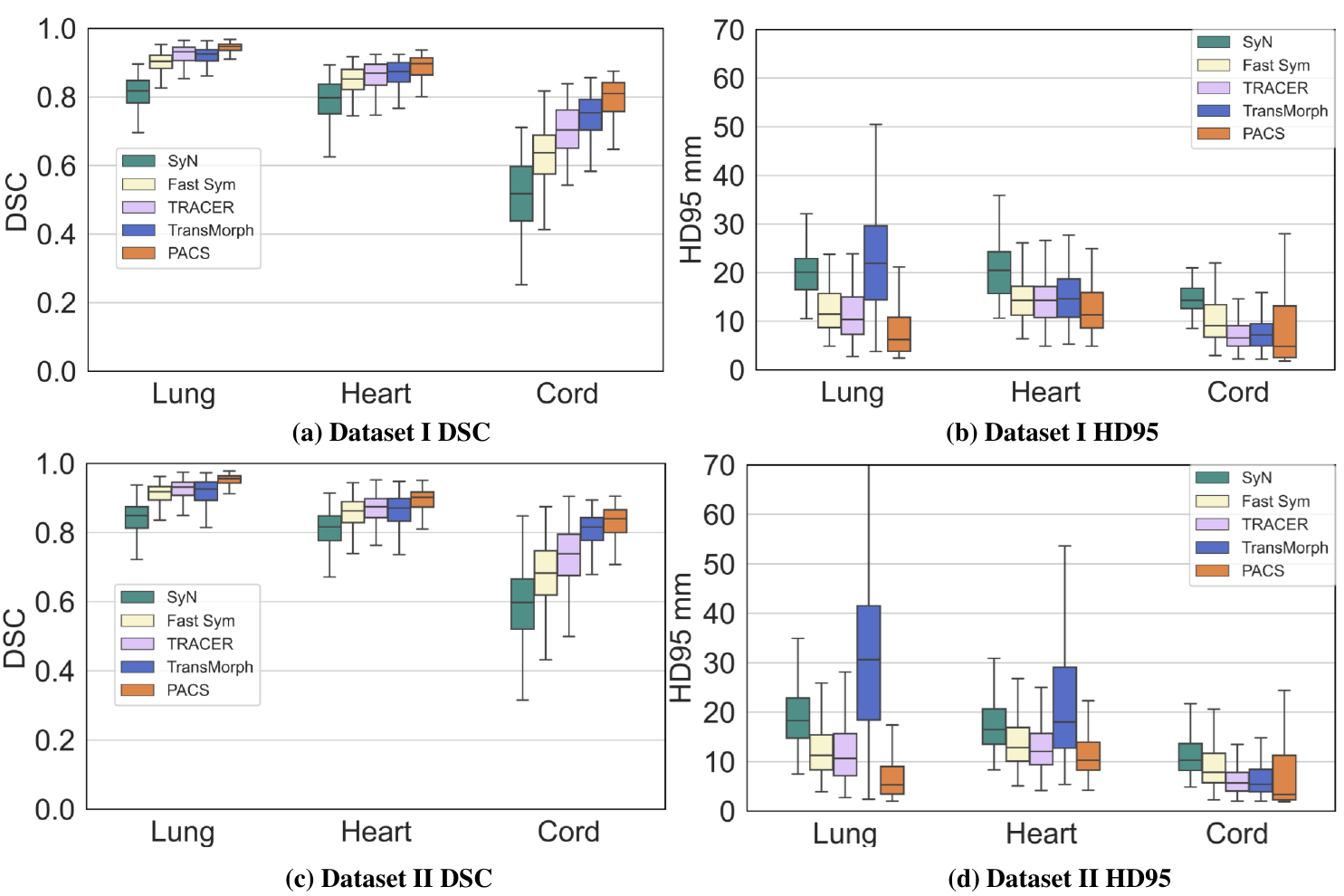}
			\vspace{-0.05cm}\setlength{\belowcaptionskip}{-0.4cm}\setlength{\abovecaptionskip}{0.08cm}\caption{\small DSC and HD95 accuracies of various methods on the testing datasets. } \label{fig:box_plot}
			
			%Labels off on Fig 3a-d - Dataset II missing on left side, dataset I missing on right side
		\end{center}
\end{figure}

\begin{figure}[ht]
		\begin{center}
			\includegraphics[width=1.0\columnwidth,scale=1]{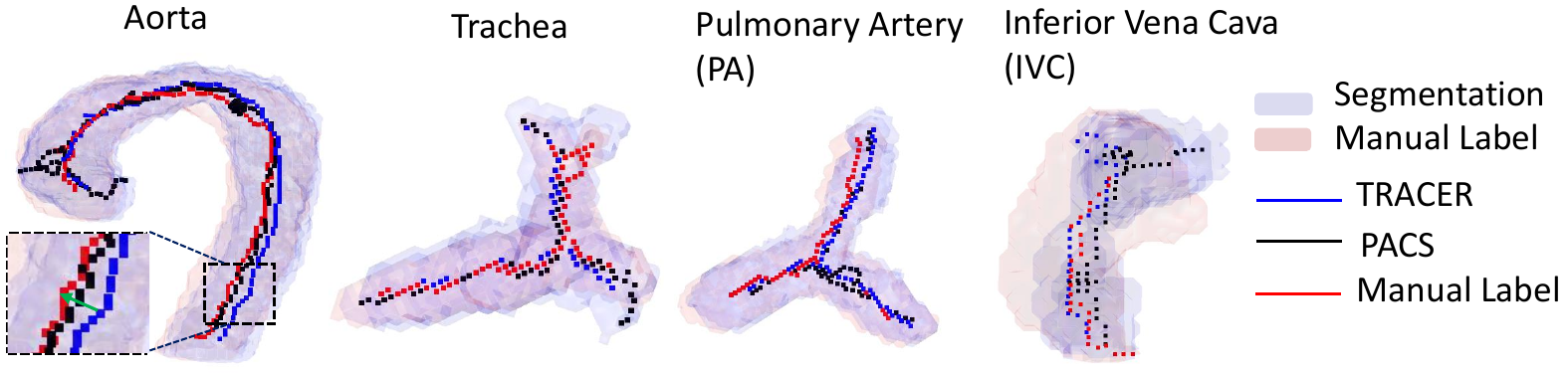}
			\vspace{-0.05cm}\setlength{\belowcaptionskip}{-0.4cm}\setlength{\abovecaptionskip}{0.08cm}\caption{\small Medial axis skeletons produced from PACS, TRACER, and manual segmentation of aorta, PA, IVC, and trachea on a sample patient.} \label{fig:medial_axis_show}
		\end{center}
\end{figure}

In terms of aligning healthy tissues, PACS was the most accurate method followed by TRACER (Figure.~\ref{fig:box_plot}). Transmorph showed wide variation in accuracy for the lungs and heart with HD95 metric. FastSym was slightly less accurate than TRACER. Improved accuracy with PACS for normal tissues was also evident when computing the MCD metric as shown in Table.~\ref{tab:reg_acc_stat}. TRACER was less accurate than PACS for all tubular structures except trachea. All DL methods resulted in a mean MCD that was within the diameter of aorta (ranging from 3.3 cm for females to 3.6 cm for males)\cite{Mao2008} and the trachea 2.3 cm on average\cite{Epstein2005}. Only PACS, TRACER and FastSym resulted in a mean MCD within the known diameter variations of the PA of 2.7 cm to 3.3 cm\cite{Edwards1998}. PACS, TRACER, and Transmorph achieved mean MCD within acceptable diameter of IVC in healthy and non-hypertensive patients (1.45 $\pm$ .16 cm)\cite{Besli2015}. Hypertensive patients have a larger diameter of IVC. SyN was the least accurate of all the methods. Fig.~\ref{fig:medial_axis_show} shows the medial axis skeleton from clinical delineation on the fixed image for a reference patient together with the medial axis skeletons produced from deformed segmentations produced using PACS and TRACER. 
\\
\begin{table*} [h]
\centering{\caption{Tumor preservation metrics computed on three testing datasets \textcolor{black}{using the metrics of $\Delta$T, M$_{lexs}$\% and MSE}. } 
\label{tab:reg_acc_stats} 
	%\centering
	%\small
	%\scriptsize
    \tiny
	%			\centering
	%\begin{tabular}{|c|c|c|c|c|c|c|c|c|c|}
        \begin{tabular}{|c|c|c|c|c|c|c|c|c|c|}
		\hline 
%		\hline 
        \multirow{1}{*}{Method} & \multicolumn{3}{c|}{Dataset I (380 pairs)} & \multicolumn{3}{c|}{Dataset II (756 pairs)} & \multicolumn{3}{c|}{Dataset III (41 pairs)} \\
       
        \cline{1-10}{} & {$\Delta T$ \% $\downarrow$} & {M$_{lexs}$\% $\downarrow$} & {\textcolor{black}{MSE $\downarrow$}} & {$\Delta T$ \% $\downarrow$} & {M$_{lexs}$\% $\downarrow$} & {\textcolor{black}{MSE $\downarrow$}} & {$\Delta T$ \% $\downarrow$} & {M$_{lexs}$\% $\downarrow$} & {\textcolor{black}{MSE $\downarrow$}}  \\ 
        
        \hline
	
            {SyN\cite{avants2008symmetric}}&{6.91$\pm$4.73}&{11.11$\pm$3.50}&{\textcolor{black}{0.14$\pm$0.076}}&{8.74$\pm$5.62}&{11.0$\pm$4.85}&{\textcolor{black}{0.064$\pm$0.017}}&{3.18$\pm$3.42} & {3.16$\pm$1.85}&{\textcolor{black}{0.040$\pm$0.048}}\\

		\hline
			
                {Fast Sym\cite{Mok2020}}&{18.87$\pm$13.67}&{14.0$\pm$5.23}&{\textcolor{black}{0.21$\pm$0.057}}&{18.19$\pm$31.03}&{15.0$\pm$7.23}&{\textcolor{black}{0.25$\pm$0.12}}& {17.36$\pm$13.69} & {3.89$\pm$2.47}&{\textcolor{black}{0.12$\pm$0.062}}\\
               
		\hline
                {Transmorph\cite{chen2022transmorph}}&{54.51$\pm$22.18}&{23.0$\pm$7.18}&{\textcolor{black}{0.52$\pm$0.093}}&{31.12$\pm$57.16}&{27.0$\pm$10.03}&{\textcolor{black}{0.45$\pm$0.15}}& {45.74$\pm$27.66} & {4.27$\pm$2.26}&{\textcolor{black}{0.37$\pm$0.19}}\\
              
		\hline
             
                {PACS\cite{jiang2022TMI}}&{96.48$\pm$23.48}&{33.13$\pm$6.27}&{\textcolor{black}{0.51$\pm$0.097}}&{58.98$\pm$92.10}&{29.0$\pm$11.33}&{\textcolor{black}{0.47$\pm$0.15}}&{97.93$\pm$96.68} & {7.36$\pm$2.45}&{\textcolor{black}{0.37$\pm$0.20}}\\
               
		\hline
                {TRACER}&{\textbf{0.24$\pm$0.49}}&{\textbf{0.8$\pm$0.09}}&{\textcolor{black}{0.005$\pm$0.003}}&{\textbf{0.40$\pm$0.82}}&{\textbf{0.45$\pm$0.12}}&{\textcolor{black}{0.005$\pm$0.006}}&{\textbf{0.13$\pm$0.15}} & {\textbf{0.87$\pm$0.14}}&{\textcolor{black}{0.003$\pm$0.002}} \\
                
		\hline
	\end{tabular}
	}
\end{table*}

\begin{table} [h]
\centering{\caption{Accuracy of aligning medial axis skeleton (MAS) measured as mean of median of closest points for tubular structures.}
\label{tab:reg_acc_stat} 
	%\centering
	%\small
	\scriptsize
	%			\centering
	\begin{tabular}{|c|c|c|c|c|} 
		\hline 
%		\hline 
        \multirow{2}{*}{Method} & \multicolumn{4}{c|}{52 pairs from different patients}  \\ 
        \cline{2-5} {} & {Aorta (cm) $\downarrow$}  & {Trachea (cm) $\downarrow$} & {PA (cm) $\downarrow$} & {IVC (cm) $\downarrow$}  \\
        \hline
		{SyN\cite{avants2008symmetric}}&{3.70$\pm$0.81}&{2.31$\pm$0.20}&{4.82$\pm$2.79}&{4.95$\pm$0.65}\\
		\hline

		%	{R2N2\cite{Sandkuhler2019}}&{2.71$\pm$0.83}&{0.50$\pm$0.28}&{3.51$\pm$2.45}&{1.93$\pm$0.64}\\
		%\hline
		%	{Voxelmorph\cite{balakrishnan2019voxelmorph}}&{2.66$\pm$0.79}&{1.89$\pm$0.21}&{3.32$\pm$2.44}&{4.89$\pm$0.66}\\
		%\hline
               {Fast Sym\cite{Mok2020}}&{2.56$\pm$0.77}&{0.38$\pm$0.18}&{3.28$\pm$2.15}&{1.78$\pm$0.54}\\
		\hline
                {Transmorph\cite{chen2022transmorph}}&{1.95$\pm$0.75}&{0.40$\pm$0.25}&{3.48$\pm$2.31}&{1.02$\pm$0.58}\\
		\hline
                %{Recursive\cite{zhao2019TweeningUnsupervised}}&{2.89}&{0.60}&{3.50}&{1.95}\\
	    	%\hline
                {PACS\cite{jiang2022TMI}}&{\textbf{1.06$\pm$0.71}}&{0.15$\pm$0.15}&{\textbf{1.98$\pm$1.64}}&{\textbf{0.65$\pm$0.45}}\\
		\hline
                {TRACER}&{1.59$\pm$0.76}&{\textbf{0.13$\pm$0.14}}&{2.49$\pm$1.93}&{0.79$\pm$0.53}\\
		\hline
	\end{tabular}
	}
\end{table}
% {figures/dvf_image.png}
\begin{figure*}[ht]
    \begin{center}
	\includegraphics[width=.9\columnwidth,scale=1]{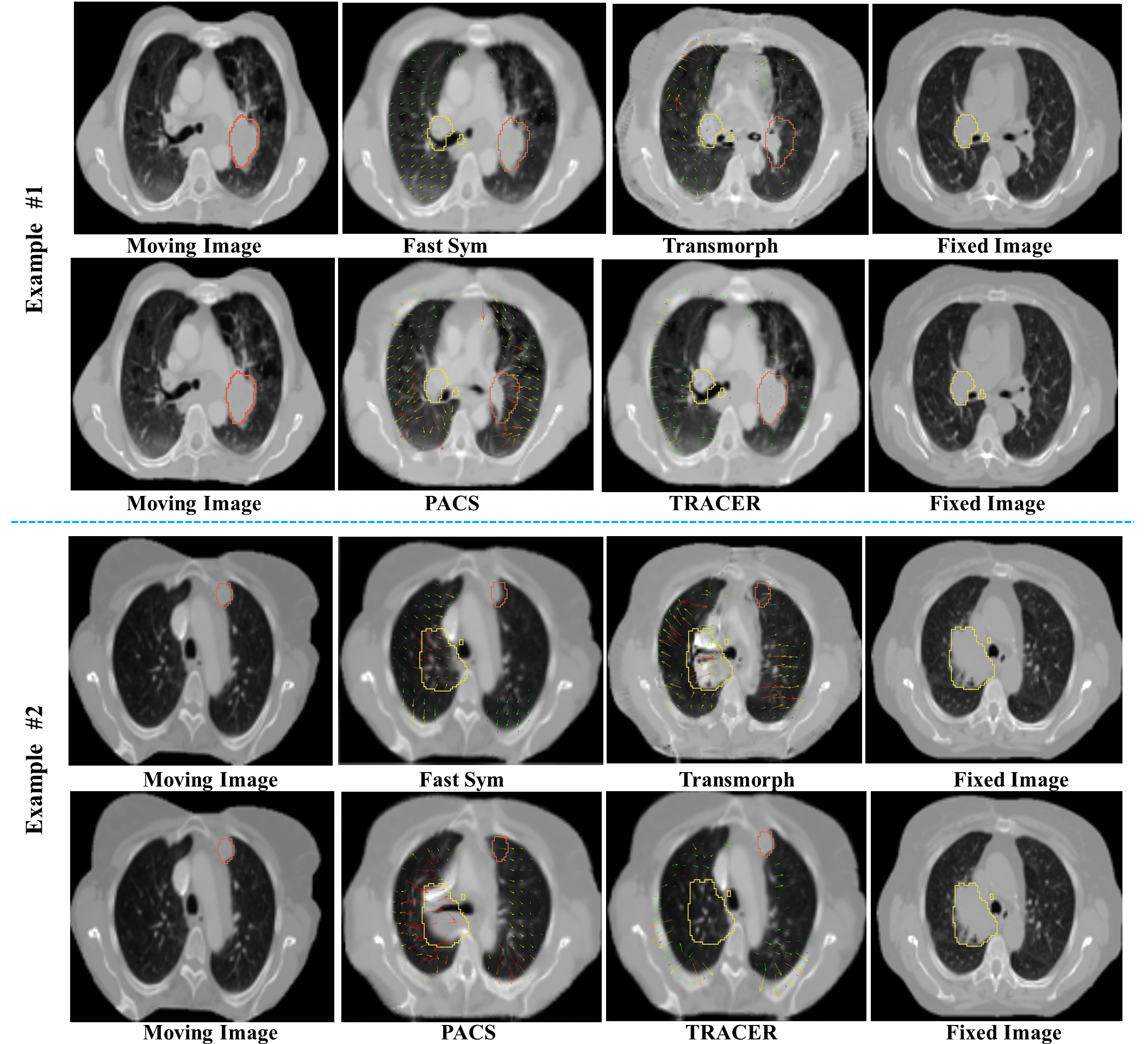}
	   \vspace{-0.00cm}\setlength{\belowcaptionskip}{-0.4cm}\setlength{\abovecaptionskip}{0.08cm}\caption{\small Registration for 2 different patient pairs containing tumor on completely different locations as well as different gender (Example $\#$2). \textcolor{black}{Red and yellow contours indicate the tumor region on moving and fixed images.}} \label{fig:seg_overlay}
    \end{center}
\end{figure*}

\subsubsection{Qualitative results}
Fig.\ref{fig:seg_overlay} shows registration results for two representative examples with varying tumor location and sizes. The second case also shows a female aligned to a male patient. As shown, TRACER best preserved the tumor occurring in the moving image and did not produce unrealistic distortions in places containing tumor on the fixed image. PACS resulted in complete erosion of the tumor in the moving image as well as unrealistic deformations in the location corresponding to tumor occurring in the fixed image. FastSym was slightly more robust to the presence of tumors but was not able to preserve the tumor on the moving image as well as TRACER. Both PACS and Transmorph produced unrealistic stretching of the right proximal bronchial tree to a considerable extent in the second patient and to a lesser degree in the first patient due a tumor on the fixed image abutting the right mediastinum. The distortion of the pulmonary artery is also visible for both patients when using PACS and Transmorph. TRACER on the other hand was unaffected by the tumor on the fixed image and preserved the tumor on the moving images. 

\subsubsection{Suitability for VBA}
\textcolor{black}{Registration quality filtering was performed using DSC metric by comparing resampled and reference lung contours (lung DSC $<$ 0.8) similar to registration filtering performed in prior work\cite{MYLONA2019}. The count of patients excluded through such filtering is shown in Table.~\ref{tab:VBA}.} Our analysis showed that both PACS and Transmorph resulted in the least number of excluded patients followed by TRACER. SyN\cite{avants2008symmetric} that is commonly used in VBA analysis resulted in an exclusion of $\geq$ 45\% of patients. FastSym was less effective than TRACER but more accurate SyN. TRACER showed similar variation in the patients that were removed for the two different references, indicating robustness to gender and lung lobes.

On the other hand, PACS resulted in the largest error in the planned dose to the tumor measured between undeformed moving and deformed moving image, indicating lack of tumor geometry preservation. FastSym and Transmorph were similar in terms of the tumor dose difference. Although SyN showed a small error in the dose to the tumor, it was ineffective in aligning most patients. TRACER resulted in the least planned tumor dose error.
\\
\begin{table} [h]
\centering{\caption{Suitability for VBA. Number of excluded patients (lung DSC $<$ 0.8) and tumor dose preservation $\Delta$PTD Gy by various methods measured using 41 cases aligned to canonical female and male reference in Dataset III.}
\label{tab:VBA} 
	%\centering
	\scriptsize
        %\small
	%			\centering
	%\begin{tabular}{|c|c|c|c|c|c|c|} 
        \begin{tabular}{|c|c|c|c|c|c|c|} 
		\hline  
        \multirow{2}{*}{Method} & \multicolumn{3}{c|}{Female reference $\downarrow$} & \multicolumn{3}{c|}{Male reference $\downarrow$}  \\ 
        \multirow{2}{*}{} & \multicolumn{2}{c}{Excluded cases} & {Tumor dose difference} & \multicolumn{2}{c}{Excluded cases} & {Tumor dose difference} \\
        \cline{2-7} {} & {Left Lung}  & {Right Lung} & {$\Delta$PTD Gy} & {Left Lung} & {Right Lung} & {$\Delta$PTD Gy} \\
        \hline
        {SyNs\cite{avants2008symmetric}}&{18}&{26}& {0.01$\pm$0.01} &{21}&{26} & {0.01$\pm$0.01}\\
	\hline
   %     {R2N2\cite{Sandkuhler2019}}&{23}&{34}&{26}&{33}\\
%	\hline
%	{Voxelmorph\cite{balakrishnan2019voxelmorph}}&{13}&{21}&{18}&{21}\\
%	\hline
        {Fast Sym\cite{Mok2020}}&{11}&{11}& {0.17$\pm$0.38} &{7}&{16}&{0.41$\pm$0.75}\\
	\hline
        {Transmorph\cite{chen2022transmorph}}&{0}&{0}&{0.14$\pm$0.14}&{1}&{2}&{0.24$\pm$0.26}\\
	\hline
        {PACS\cite{jiang2022TMI}}&{0}&{0}& {28.45$\pm$15.30}&{1}&{2}&{22.92$\pm$19.91}\\
	\hline
        {TRACER}&{5}&{5}&{0.01$\pm$0.02}&{6}&{5}& {0.013$\pm$0.027}\\
	\hline
	\end{tabular}
	}
\end{table}

\subsubsection{Ablation experiments}
\begin{figure*}
    \begin{center}
	\includegraphics[width=1.0\columnwidth,scale=1]{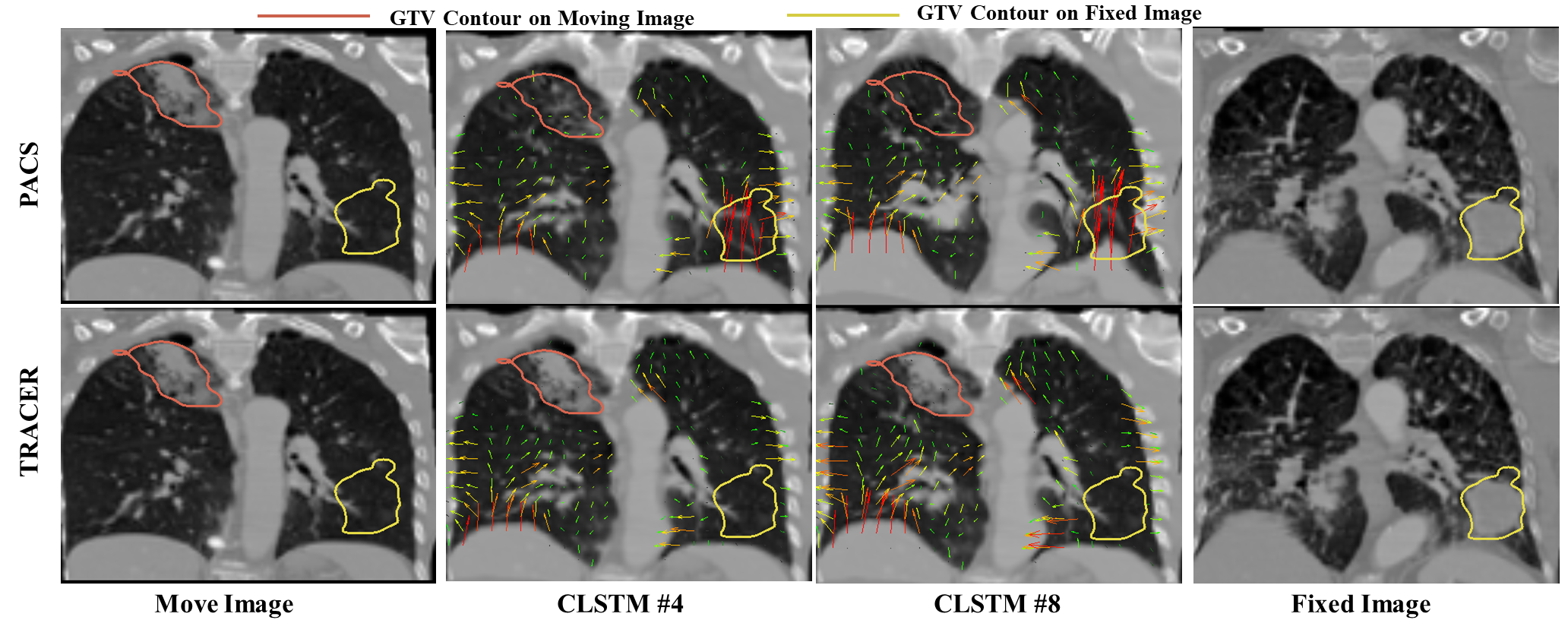}
	   \vspace{-0.05cm}\setlength{\belowcaptionskip}{-0.4cm}\setlength{\abovecaptionskip}{0.08cm}\caption{\small TRACER used to align a representative pair of images containing large tumors on diferent lobes of the lungs. Deformation map (top row) and resampled moving images (bottom row) produced in CLSTM steps 2,4,6, and 8 (b) without tumor preservation (upper) and with tumor preservation constraints (below) are also shown.} \label{fig:dev_overlay}
    \end{center}
\end{figure*}

Table~\ref{tab:ablation} shows the tumor preservation and tissue segmentation accuracies with and without CLSTM and with and without tumor conditioning constraints. Removal of tumor conditioning drastically reduced the tumor preservation accuracy compared to when using these constraints. Tumor conditioning in the forward direction was more important for tumor preservation than in the inverse direction. Conversely, tumor conditioning reduced the segmentation accuracy for the healthy tissues, which was slightly improved when using CLSTM. On the other hand, CLSTM without tumor conditioning reduced tumor preservation accuracy furthermore as the model tries to progressively shrink the tumor voxels to better align with the healthy tissue occurring in the same anatomic location. Fig.~\ref{fig:dev_overlay} shows an example of progressively refined alignment without and with tumor conditioning through the different CLSTM steps. As seen, without the CLSTM, the tumor geometry is progressively lost in the former case but is preserved in the latter case. On the other hand, when not using tumor conditioning, the chest wall is pulled towards lungs to match with the tumor occurring in the fixed image, thus resulting in physically unrealistic deformation compared to when using the tumor conditioning with CLSTM. The successive deformation maps are also shown alongside the deformed images. 

\begin{figure*}[t]
		\begin{center}
			\includegraphics[width=1\columnwidth,scale=1]{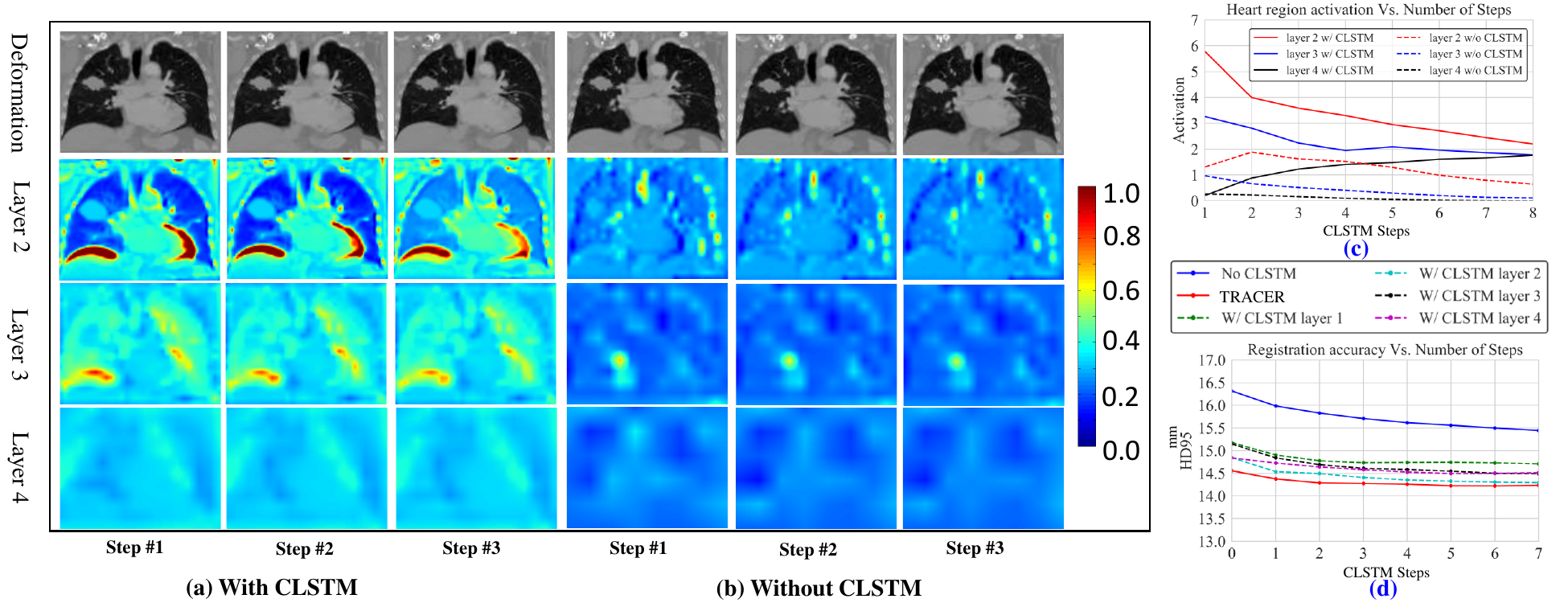}
			\vspace{-0.05cm}\setlength{\belowcaptionskip}{-0.4cm}\setlength{\abovecaptionskip}{0.08cm}\caption{\small {(b) Feature activations produced in CNN layers 2, 3, and 4 for steps 1, 2, 3 without CLSTM, and (a) with CLSTM. (c) shows mean feature activations within heart in the layers 2, 3, and 4. (d) shows DSC accuracy with increasing number of CLSTM steps.} } \label{fig:sCLSTM_show}
		\end{center}
\end{figure*}

\begin{table} [ht]
\centering{\caption{Ablation testing. Forward conditioning uses moving image, inverse uses fixed image, and both uses both images.}
\label{tab:ablation} 
	%\centering
	\scriptsize
        %\small
	%			\centering
	\begin{tabular}{|c|c|c|c|c|c|c|} 
		\hline 
%		\hline 
		{CLSTM}&{Condition}&{$\Delta$T \% $\downarrow$}&{Lung $\uparrow$}&{Heart $\uparrow$}&{Cord $\uparrow$}\\
		\hline
 
 {$\times$} &{None}&{80.2$\pm$24.1} &{0.94$\pm$0.02}& {0.86$\pm$0.05}&{0.75$\pm$0.09} \\% {No Bi-rigidity}& {7.24}\\ 
  {$\times$} & {Inverse}& {30.8$\pm$18.9}&{0.93$\pm$0.02} &{0.85$\pm$0.05}&{0.69$\pm$0.09} \\% {No Bi-
    {$\times$}& {Forward}&{0.83$\pm$1.56} &{0.93$\pm$0.02}& {0.85$\pm$0.05}&{0.69$\pm$0.09} \\% {No Bi-
   	{$\times$} & {Both}&{0.54$\pm$0.67} &{0.90$\pm$0.06}& {0.85$\pm$0.06}&{0.68$\pm$0.10} \\% {No
    
\hline 
  {$\checkmark$} &{None}&{96.5$\pm$23.5} &{0.94$\pm$0.02}& {0.87$\pm$0.05}&{0.76$\pm$0.09} \\% {No Bi-rigidity}& {7.24}\\ 
  {$\checkmark$} &{Inverse}&{41.9$\pm$46.7} &{0.93$\pm$0.03}& {0.85$\pm$0.05}&{0.72$\pm$0.09} \\% {No Bi-
    {$\checkmark$} &{Forward}&{0.31$\pm$0.61} &{0.93$\pm$0.03}& {0.85$\pm$0.05}&{0.72$\pm$0.10} \\% {No Bi-
   	{$\checkmark$} & {Both}& {0.24$\pm$0.49}	&{0.92$\pm$0.03}&{0.85$\pm$0.06}&{0.70$\pm$0.09}\\% {TRACER} & {7.53}\\
    \hline 
	\end{tabular}
	}
\end{table}
Next, we studied the importance of using a stacked CLSTM on the registration accuracy. As shown in Fig.~\ref{fig:sCLSTM_show}, the feature activations are more pronounced towards voxels with large differences between the two patient scans such as the heart and the tumor when using the CLSTM. Furthermore, different areas of the image show higher feature activations for different CLSTM steps, which allows TRACER to capture deformations at varying scales and locations. Feature activations across the encoder layers follow different trends across the steps when using CLSTM with the activations decreasing drastically in the later steps for layer 1 and increasing for layer 3 as shown in Fig.~\ref{fig:sCLSTM_show} C. The feature activations are relatively low and follow the same trend when not using the stacked CLSTM. Also, the registration accuracy increases with the number of stacked CLSTMs (Fig.~\ref{fig:sCLSTM_show} D). 

\section{Discussion}
We introduced TRACER, an inter-patient DL-DIR method for aligning thoracic CTs of patients with LA-NSCLC. TRACER was the most accurate approach for preserving conditioned structures such as the tumors and prevented physically unrealistic anatomy deformations in locations where new tumors occurred on fixed image. TRACER was second most accurate for aligning organs and tubular organs after PACS\cite{jiang2022CBCTTMI}. PACS on the other hand was ineffective to preserve tumors. Unlike prior approaches using rigidity constraints\cite{MYLONA2019,staring2007}, TRACER is fully automated as it does not require manual delineations for conditioned structures and can handle tumors occurring at non-corresponding locations in moving and fixed images. 

VBA methods have primarily utilized iterative non deep learning methods including rigid registration\cite{Beaumont2019} and elastic DIR\cite{Craddock2023,MYLONA2019, palma2016}. We, for the first time analyzed the feasibility of using DL-DIR methods for addressing inter-patient tumor variabilities for VBA. Our analysis showed that TRACER as well as other DL-DIR methods were more accurate than iterative DIR using SyN\cite{avants2008symmetric}. Despite preserving tumor geometry, SyN resulted in the largest number of excluded examples due to poor registration. TRACER on the other hand was resilient to anatomic differences even in the presence of non-corresponding tumors and patient gender as evidenced by relatively fewer excluded cases with highest tumor geometry preservation. Of note, we used lung registration accuracy as a metric to filter patients, but structures can also be used. TRACER was resistant to the presence of tumors abutting mediastinum and better preserved geometry compared with PACS and Transmorph. 

%Although inter-patient registration is a well researched topic in the context of atlas based registration in radiation oncology applications, the driving motivation for these works is normal tissue segmentation\cite{haq2019,xu2019deepatlas,dalca2019unsupervised,liu2021same}. We, for the first time studied inter-patient registration in the context of aligning patients with visible tumors in the lung. 
%\\

TRACER shares the same registration network architecture as PACS\cite{jiang2022CBCTTMI}. TRACER does not include the segmentation subnetwork used in PACS, and makes use of explicit tumor conditioning as well as bidirectional tumor incompressibility constraints. Our results showed improved tumor preservation when combining tumor conditioning with the stacked 3DCLSTM approach. In particular, removal of tumor conditioning resulted in a dramatic tumor mass loss, thus suggesting that conditioning is critical for tumor preservation and removing physically unrealistic deformations. A key advantage of the stacked CLSTM architecture is that it allows to capture varying amounts of local tissue deformation in the different layers. In addition, CLSTM natually allows to compute spatially and temporally varying velocity field, which is not possible with single step registration methods\cite{balakrishnan2019voxelmorph,Mok2020}. It is known that the ability to model spatially and temporally varying velocity fields reduces chances of high energy transformations in the presence of substantial anatomic differences\cite{Ashburner2007}. 

As limitation, the current work was focused on developing and evaluating a inter-patient registration method and hence did not perform an indepth study of downstream effects on VBA itself, which is for future studies. In summary, TRACER generated tumor preserving accurate registration of inter-patient thoracic CTs. 

\section{Conclusion}
We introduced TRACER, an inter-patient registration method for aligning thoracic CTs of patients with LA-NSCLC. TRACER was the most accurate approach for preserving conditioned structures such as the tumors and prevented physically unrealistic anatomy deformations in locations where new tumors occurred on fixed image. TRACER also showed capability to balance registration accuracy for normal tissues while preserving tumor geometry, thus indicating its suitability for voxel-based analysis. \textcolor{black}{Further studies are required to evaluate its utility in population level outcomes modeling studies using voxel-based analysis.}

\section{Acknowledgments}
This work was partially supported by NCI R01CA25881 and the MSK Cancer Center support grant/core grant P30 CA008748, and the Korea Health Technology R$\&$D Project through the Korea Health Industry Development Institute (KHIDI), funded by the Ministry of Health $\&$ Welfare, Republic of Korea (HI19C1234).

\section*{References}
\bibliographystyle{splncs04.bst}
\bibliography{mybibliography}

\begin{thebibliography}{10}
\providecommand{\url}[1]{\texttt{#1}}
\providecommand{\urlprefix}{URL }
\providecommand{\doi}[1]{https://doi.org/#1}

\bibitem{aerts2015data}
Data from \textsc{NSCLC}-radiomics. \textsc{T}he \textsc{C}ancer
  \textsc{I}maging \textsc{A}rchive (2015)

\bibitem{Kim2021CycleMorph}
Cyclemorph: Cycle consistent unsupervised deformable image registration.
  Medical Image Analysis  \textbf{71},  102036 (2021)

\bibitem{Ahmad2019}
Ahmad, S., Fan, J., Dong, P., Cao, X., Yap, P.T., Shen, D.: Deep learning
  deformation initialization for rapid groupwise registration of inhomogeneous
  image populations. Front Neuroinform  \textbf{34}(13) (2019). \doi{doi:
  10.3389/fninf.2019.00034}

\bibitem{ALAM2021883}
Alam, S.R., Zhang, P., Zhang, S.Y., Chen, I., Rimner, A., Tyagi, N., Hu, Y.C.,
  Lu, W., Yorke, E.D., Deasy, J.O., Thor, M.: Early prediction of acute
  esophagitis for adaptive radiation therapy. Int J Radiat Oncol Biol Phys
  \textbf{110}(3),  883--892 (2021)

\bibitem{Ashburner2007}
Ashburner, J.: A fast diffeomorphic image registration algorithm. Neuroimage
  \textbf{38}(1),  95--113

\bibitem{avants2008symmetric}
Avants, B.B., Epstein, C.L., Grossman, M., Gee, J.C.: Symmetric diffeomorphic
  image registration with cross-correlation: evaluating automated labeling of
  elderly and neurodegenerative brain. Medical image analysis  \textbf{12}(1),
  26--41 (2008)

\bibitem{balakrishnan2019voxelmorph}
Balakrishnan, G., Zhao, A., Sabuncu, M.R., Guttag, J., Dalca, A.V.: Voxelmorph:
  a learning framework for deformable medical image registration. IEEE
  transactions on medical imaging  \textbf{38}(8),  1788--1800 (2019)

\bibitem{Beasley2019}
Beasley, W., Thor, M., McWilliam, A., Green, A., R, M., Slevin, N., Olsson, C.,
  Pettersson, N., Finizia, C., Estilo, C., Riaz, N., Lee, N., Deasy, J., van
  Herk, M.: Image-based data mining to probe dosimetric correlates of
  radiation-induced trismus. Int J Radiat Oncol Biol Phys  \textbf{102}(4),
  1330--1338 (2019)

\bibitem{Beaumont2019}
Beaumont, J., Acosta, O., Devillers, A., Palard-Nevello, X., Chajon, E.,
  Crevoisier, R.d., Castelli, J.: Voxel-based identification of local
  recurrence sub-regions from pre-treatment pet/ct for locally advanced head
  and neck cancers. EJNMMI Res  \textbf{9}(90) (2019)

\bibitem{Besli2015}
Besli, F., Kecebas, M., Caliskan, S., Dereli, S., Baran, I., Turker, Y.: The
  utility of inferior vena cava diameter and the degree of inspiratory collapse
  in patients with systolic heart failure. Am J Emer Med  \textbf{33}(5),
  653--7

\bibitem{chen2022transmorph}
Chen, J., Frey, E.C., He, Y., Segars, W.P., Li, Y., Du, Y.: Transmorph:
  Transformer for unsupervised medical image registration. Medical image
  analysis  \textbf{82},  102615 (2022)

\bibitem{Craddock2023}
Craddock, M., Nestle, U., Koenig, J., Schimek-Jasch, T., Kremp, S., Lenz, S.,
  Banfill, K., Davey, A., Price, G., Salem, A., Faivre-Finn, C., van Herk, M.,
  McWilliam, A.: Cardiac function modifies the impact of heart base dose on
  survival: A voxel-wise analysis of patients with lung cancer from the
  \textsc{PET}-plan trial. J Thorac Oncol  \textbf{18}(1),  57--66 (2023)

\bibitem{dalca2019unsupervised}
Dalca, A.V., Balakrishnan, G., Guttag, J., Sabuncu, M.R.: Unsupervised learning
  of probabilistic diffeomorphic registration for images and surfaces. Medical
  Image Anal.  \textbf{57},  226--236 (2019)

\bibitem{Edwards1998}
Edwards, P., Bull, R., Coulden, R.: \textsc{CT} measurement of main pulmonary
  artery diameter. Br J Radiol  \textbf{71}(850),  1018--20 (1998)

\bibitem{Epstein2005}
Epstein, S.: Anatomy and physiology of tracheostomy. Respir Care
  \textbf{50}(4),  476--82 (2005)

\bibitem{haq2019}
Haq, R., Berry, S.L., Deasy, J., Hunt, M., Veeraraghavan, H.: Dynamic
  multiatlas selection-based consensus segmentation of head and neck structures
  from ct images. Med Phys  \textbf{46}(12),  5612--5622 (2019)

\bibitem{jaderberg2015spatial}
Jaderberg, M., Simonyan, K., Zisserman, A., Kavukcuoglu, K.: Spatial
  transformer networks. arXiv preprint arXiv:1506.02025  (2015)

\bibitem{jiang2022CBCTTMI}
Jiang, J., Veeraraghavan, H.: One shot \textsc{PACS}: \textsc{P}atient specific
  anatomic context and shape prior aware recurrent registration-segmentation of
  longitudinal thoracic cone beam \textsc{CT}s. IEEE Trans Med Imaging  (2022).
  \doi{doi: 10.1109/TMI.2022.3154934}

\bibitem{jiang2022TMI}
Jiang, J., Veeraraghavan, H.: One shot \textsc{PACS}: \textsc{P}atient specific
  anatomic context and shape prior aware recurrent registration-segmentation of
  longitudinal thoracic cone beam \textsc{CT}s. IEEE Trans Med Imaging p.
  PP:10.1109/TMI.2022.3154934 (2022)

\bibitem{kalendralis2020fair}
Kalendralis, P., Shi, Z., Traverso, A., Choudhury, A., Sloep, M., Zhovannik,
  I., Starmans, M.P., Grittner, D., Feltens, P., Monshouwer, R., et~al.:
  \textsc{FAIR}-compliant clinical, radiomics and \textsc{DICOM} metadata of
  \textsc{RIDER}, interobserver, lung1 and head-neck1 \textsc{TCIA}
  collections. Med Phys  \textbf{47}(11),  5931--5940 (2020)

\bibitem{Konig2016RadOnc}
König, L., Derksen, A., Papenberg, N., Haas, B.: Deformable image registration
  for adaptive radiotherapy with guaranteed local rigidity constraints. Radiat
  Oncol  \textbf{11}(122) (2016)

\bibitem{liu2021same}
Liu, F., Yan, K., Harrison, A.P., Guo, D., Lu, L., Yuille, A.L., Huang, L.,
  Xie, G., Xiao, J., Ye, X., et~al.: \textsc{SAME}: Deformable image
  registration based on self-supervised anatomical embeddings. In: Medical
  Image Computing and Computer-Assisted Intervention. pp. 87--97 (2021)

\bibitem{Mao2008}
Mao, S., Ahmadi, N., Shah, B., Beckmann, D., Chen, A., Ngo, L., Flores, F.,
  Gao, Y., Budoff, M.: Normal thoracic aorta diameter on cardiac computed
  tomography in healthy asymptomatic adult; impact of age and gender. Acad
  Radiol  \textbf{15}(7),  827--834 (2008)

\bibitem{Mok2020}
Mok, T.C.W., Chung, A.C.S.: Large deformation diffeomorphic image registration
  with laplacian pyramid networks. In: Medical Image Computing and Computer
  Assisted Intervention. pp. 211--221. Springer International Publishing (2020)

\bibitem{MokChung2021MICCAI}
Mok, T.C.W., Chung, A.C.S.: Conditional deformable image registration with
  convolutional neural network. In: Medical Image Computing and Computer
  Assisted Intervention -- MICCAI 2021. pp. 35--45. Springer International
  Publishing (2021)

\bibitem{monti2018}
Monti, S., Pacelli, R., Cella, L., Palma, G.: Inter-patient image registration
  algorithms to disentangle regional dose bioeffects. Sci Rep  \textbf{8},
  ~4915 (2018)

\bibitem{MYLONA2019}
Mylona, E., Acosta, O., Lizee, T., Lafond, C., Crehange, G., Magné, N.,
  Chiavassa, S., Supiot, S., {Ospina Arango}, J.D., Campillo-Gimenez, B.,
  Castelli, J., {de Crevoisier}, R.: Voxel-based analysis for identification of
  urethrovesical subregions predicting urinary toxicity after prostate cancer
  radiation therapy. Int J Radiat Oncol Biol Phys  \textbf{104}(2),  343--354
  (2019)

\bibitem{niedzielski2016}
Niedzielski, J., Yang, J., Stingo, F., Martel, M., Mohan, R., Gomez, D.,
  Briere, T., Liao, Z., Court, L.: Objectively quantifying radiation
  esophagitis with novel computed tomography-based metrics. Intl J Radiat Oncol
  Bio Physics  \textbf{94}(2),  385--93 (2016)

\bibitem{palma2016}
Palma, G., Monti, S., D'Avino, V., Conson, M., Liuzzi, R., Pressello, M.,
  Donato, V., Deasy, J., Quarantelli, M., Pacelli, R., Cella, L.: A voxel-based
  approach to explore local dose differences associated with radiation-induced
  lung damage. Int J. Radiat Oncol Biol Phys  \textbf{96}(1),  127--33 (2016)

\bibitem{Sandkuhler2019}
Sandk{\"u}hler, R., Andermatt, S., Bauman, G., Nyilas, S., Jud, C., Cattin,
  P.C.: Recurrent registration neural networks for deformable image
  registration. Adv. Neural Inf. Process. Syst.  \textbf{32} (2019)

\bibitem{staring2007}
Staring, M., Klein, S., Pluim, J.: A rigidity penalty term for nonrigid
  registration. Medical Physics  \textbf{34}(11),  4098--108 (2007)

\bibitem{um2020multiple}
Um, H., Jiang, J., Thor, M., Rimner, A., Luo, L., Deasy, J.O., Veeraraghavan,
  H.: Multiple resolution residual network for automatic thoracic
  organs-at-risk segmentation from ct (2020)

\bibitem{xu2019deepatlas}
Xu, Z., Niethammer, M.: Deepatlas: Joint semi-supervised learning of image
  registration and segmentation. In: MICCAI. pp. 420--429. Springer (2019)

\bibitem{yang2018conditional}
Yang, Y., Soatto, S.: Conditional prior networks for optical flow. In:
  Proceedings of the European Conference on Computer Vision (ECCV). pp.
  271--287 (2018)

\bibitem{ZhangInverseConsistent2018}
Zhang, J.: Inverse-consistent deep networks for unsupervised deformable image
  registration. CoRR  \textbf{abs/1809.03443} (2018)

\bibitem{Zhang2021}
Zhang, Y., Wu, X., Gach, H., Li, H., Yang, D.: Groupregnet: a groupwise
  one-shot deep learning-based 4d image registration method. Phys Med Biol
  \textbf{66}(4),  045030

\end{thebibliography}

\end{document}